\documentclass[sigconf]{acmart} %sigconf, anonymous

\renewcommand\footnotetextcopyrightpermission[1]{} % removes footnote with conference information in first column
\usepackage[english]{babel}
\usepackage[utf8x]{inputenc}
\usepackage[T1]{fontenc}
\usepackage{ucs}

\usepackage{amsmath}
\usepackage{amsmath}
\usepackage{amsfonts}
\usepackage{graphicx}
\usepackage{tikz}
\usetikzlibrary{shapes,arrows,chains,calc,bending}

\usepackage[]{hyperref}%[colorlinks=true, allcolors=blue] draft
\usepackage[neveradjust]{paralist}
\usepackage{multirow}

\usepackage{microtype}

\usepackage{booktabs} % For formal tables
\usepackage{arydshln}

\colorlet{lcfree}{green}
\colorlet{lcnorm}{blue}
\colorlet{lccong}{red}
\pgfdeclarelayer{marx}
\pgfsetlayers{main,marx}
\providecommand{\cmark}[2][]{%
	\begin{pgfonlayer}{marx}
		\node [nmark] at (c#2#1) {#2};
	\end{pgfonlayer}{marx}
} 
\providecommand{\cmark}[2][]{\relax} 
\copyrightyear{2021}
\acmYear{2021}
\begin{document}
	
	\title{Architecture and Its Vulnerabilities in Smart-Lighting Systems}%[Short title]
	
	\author{Florian Hofer}
	\orcid{0000-0001-8336-0834}
	\affiliation{%
		\institution{Free University of Bolzano-Bozen}
		\streetaddress{Piazza Domenicani 3}
		\city{Bolzano}
		\state{}
		\postcode{39100, IT}
	}
	\email{florian.hofer@stud-inf.unibz.it}
	
	\author{Barbara Russo}
	\orcid{0000-0003-3737-9264}
	\affiliation{%
		\institution{Free University of Bolzano-Bozen}
		\streetaddress{Piazza Domenicani 3}
		\city{Bolzano}
		\state{}
		\postcode{39100, IT}
	}
	\email{barbara.russo@unibz.it}
	
	\renewcommand{\shortauthors}{F. Hofer and B. Russo}
	
	%	\begin{CCSXML}
	%		<ccs2012>
	%		<concept>
	%		<concept_id>10010520.10010553</concept_id>
	%		<concept_desc>Computer systems organization~Embedded and cyber-physical systems</concept_desc>
	%		<concept_significance>500</concept_significance>
	%		</concept>
	%		<concept>
	%		<concept_id>10010520.10010521.10010537</concept_id>
	%		<concept_desc>Computer systems organization~Distributed architectures</concept_desc>
	%		<concept_significance>500</concept_significance>
	%		</concept>
	%		<concept>
	%		<concept_id>10002978.10003014</concept_id>
	%		<concept_desc>Security and privacy~Network security</concept_desc>
	%		<concept_significance>300</concept_significance>
	%		</concept>
	%		</ccs2012>
	%	\end{CCSXML}
	%	
	%	\ccsdesc[500]{Computer systems organization~Embedded and cyber-physical systems}
	%	\ccsdesc[500]{Computer systems organization~Distributed architectures}
	%	\ccsdesc[300]{Security and privacy~Network security}	
	
	
	\begin{abstract}		
		Industry 4.0 embodies one of the significant technological changes of this decade. Cyber-physical systems and the Internet Of Things are two central technologies in this change that embed or connect with sensors and actuators and interact with the physical environment. However, such systems-of-systems undergo additional restrictions in an endeavor to maintain reliability and security when building and interconnecting components to a heterogeneous, multi-domain  \textit{Smart-*} systems architecture. This paper presents an application-specific, layer-based approach to an offline security analysis inspired by design science that merges preceding expertise from relevant domains. With the example of a Smart-lighting system, we create a dedicated unified taxonomy for the use case and analyze its distributed Smart-* architecture by multiple layer-based models. We derive potential attacks from the system specifications in an iterative and incremental process and discuss resulting threats and vulnerabilities. Finally, we suggest immediate countermeasures for the latter potential multiple-domain security concerns.	
	\end{abstract}
	
	\maketitle
	%	\linenumbers % Line-number command, turn on line numbering.
	
	\section{Introduction}
	
	Progressive computerization brings technology into every corner and improves the automation and performance of environmental and manufacturing processes. The German Government envisions the fourth industrial revolution as an inevitable prospect for future development. This revolution endows modern systems with ``Smart'' attributes to increase operational efficiency, share information, and improve their services' quality~\cite{Leeetal2015}. These endowments allow the creation of fully flexible production systems. They bring in new business models, services, and products via \textit{Smart-*} systems such as \textit{Smart-Home} or \textit{Smart-City}~\cite{Lu2017}.
	
	Smart technologies rely on Cyber-physical systems (CPS) and the Internet of Things (IoT) to achieve such goals. \textit{Smart-*} systems operate via an autonomous, decentralized decision-making process that allows for local and faster reaction and thus enables higher efficiency and production quality~\cite{Lezzietal2018}. They run on a network of intelligent devices of diverse make and function, requiring standardized interfacing and communication. This heterogeneity could lead to inconsistencies making a system vulnerable. System attacks can exploit those vulnerabilities to eavesdrop or harm an asset's value, causing virtual and physical loss. This distress is particularly the case of \textit{Smart-Lighting} systems, where publicly installed devices may be subject to physical and cyber-attacks~\cite{AshibaniMahmoud2017}.
	
	Consequently, modern architectural designs needs must include security (\textit{security by design}). However, this is a hard-to-achieve task in a multi-domain environment where definitions and analysis models defer between the application-relevant functional models. Furthermore, a recent systematic mapping study identifies a lack of research on security for Industry 4.0 architectures~\cite{Hofer2018}.  As a result, security often experienced neglect, and existing architectural models frequently suffer from simplifications and assumptions from the ``offline''-secure world. Thus, there is a need for guidelines to build architectures and their models that incorporate security concerns. Such guidelines would also help assess systems' multi-domain vulnerabilities and propose strategic and preventive countermeasures~\cite{Lezzietal2018} or determine corrective mitigation measures that could reduce or eliminate a vulnerability~\cite{Moteff2005}.
	
	This study presents a layer-based analysis and classification technique of architectural vulnerabilities for multi-domain systems. While specific new weaknesses may appear when interconnecting such heterogeneous systems, this paper focuses on a technique that extends our knowledge of a system through existing results from relevant domains. We explore security concerns within several reference models designed from connected CPSs or IoT sub-domains to extend our assessment beyond a single-domain analysis systematically. As a practical example of such a process, we examine the decentralized \textit{Smart-Lighting} architecture of an in-field case study. Overall, the contributions of this work are:
	\begin{itemize}
		\item A technique to associate and classify architectural layers of existing reference models for CPS and IoT environments;
		\item A technique to consolidate the taxonomy of threats and attacks for vulnerability analysis of multi-domain, multi-layer Smart-* architectures;
		\item An architectural analysis and multi-domain taxonomy on vulnerabilities and attacks of a Smart-Lighting system.
	\end{itemize}
	
	We organized the rest of the paper as follows. Section~\ref{sec:relwork} and~\ref{sec:method} present related work and our methodology and evaluation strategy. In Section~\ref{sec:identification}, we examine the case study and create its multi-domain architecture reference and security taxonomy tables. In Section~\ref{sec:classification}, we apply our iterative classification technique using these tables continued by discussing threats, vulnerabilities, and possible countermeasures. Finally, we conclude in Section~\ref{sec:concl}.
	
	\section{Related work}
	\label{sec:relwork}
	
	We identified three major security topics: assessment through architecture layers, (traditional) offline analysis tools, and architecture design and patterns. In addition, we select cornerstone studies and describe their relevance in Industry 4.0 in the following.
	
	\paragraph*{Security and layers}
	Lezzi et al.~\cite{Lezzietal2018} analyze how research deals with the current cybersecurity issues in Industry 4.0 contexts, laying down the state of development regarding Smart-* architectures. The authors argue that an ideal design and development strategy considers cybersecurity from the start. The study identifies norms and guidelines for architecture security. It proposes structured solution approaches along with the taxonomy of standard cybersecurity terms. Within their list of threat identification methods, they mention a three-layer-based attack assessment technique. While they do not discuss the method's efficiency further, their concluding remarks highlight the lack of an all-layer cybersecurity analysis. 
	
	Although little research exists on vulnerability classifications in these new Smart contexts, we can adopt some published results on CPS architectures. Ashibani and Mahmoud~\cite{AshibaniMahmoud2017} redacted a generic security analysis comparing CPS technologies to traditional IT security. The article is among the first to discuss the analysis and detection of multi-layer security requirements, possible attacks, and issues for information security on three architectural layers. However, their theoretical considerations appear limited to their feasibility, and many discussed terms had non-traceable sources. 
	
	Varga \textit{et al.}~\cite{Vargaetal2017} created an analogous, IoT focused overview. The study targets the automation domain with a fourth architecture layer for data processing, highlighting security threats and mitigation. The paper displays how similar analyses can impact results from their biased viewpoint. While IoT and CPS security present similarities, the article disregards typical distributed control and treats issues as binary problems making the analysis incomplete. However, the strong data-centric viewpoint helps in the assessment of data processing systems. 
	
	Han \textit{et al.}~\cite{Hanetal2014} submit in their layer analysis a different aspect to vulnerabilities by classifying them as internal or external. They propose a four-plus-one layer architecture and a framework for an intrusion detection system (IDS). Due to the lack of a unique definition of CPS, the authors suggest an iterative application of appropriate mitigation strategies. Unfortunately, they apply this iterative notion to IDS design only. Furthermore, even though they deliver a control-centered selection of attacks for each layer, the article also admits definition issues.
	
	\paragraph*{(Traditional) offline analysis tools for security and safety} 
	Safety and security relied on design time offline analysis tools for many years, a tradition that did not change much for cybersecurity. Bolbot et al.~\cite{Bolbotetal2019} describe the relationship between the two as a conditional dependence. Their article focuses on design-time safety assurance methods, their modifications, and their integration. They identify sources of CPSs' complexity and test offline assessment techniques against them. Within the remarks of this investigation, we find the need for a systematic method for issue identification. They highlight the importance of mixing and adapting existing techniques to deal with CPS's complexities to tackle cybersecurity issues.
	
	Subramanian and Zalewski propose in \cite{SubramanianZalewski2016}, and \cite{SubramanianZalewski2018} an alternative assessment approach for non-functional requirements to connect security and safety in the CPS domain. The non-functional domain's well-defined ontology allows for an inter-dependency graph, which then propagates information as needed. The method shows how the dependencies of a single requirement can change an issue's weight. Majed et al.~\cite{Majedetal2014} suggests a framework for evaluating security exposure by weight on a connected graph. Via the shortest path, we can then identify the most accessible vulnerability. Although both are exciting approaches, the distribution of weight and path for each node remains unclear.	
	
	\paragraph*{Architecture design and patterns}	
	Alguliyev \textit{et al.}~\cite{Alguliyevetal2018} analyze and classify in a recent literature review existing research on CPS security using the CIARR model, a variant of the CIAA security requirements. This variant separates availability into resilience and reliability, suggesting that CPS's non-functional requirements vary from traditional IT. The analysis discusses approaches of architectural design to improve system security. It draws up the context and risks, offers a generalized attack tree, proposes mitigation strategies, and informs about found countermeasures and dominant future research areas. 
	
	Ryoo \textit{et al. 2015}~\cite{Ryooetal2015} try to break assessment conventions by proposing a generic new three-stage approach. The three phases collect information based on tactics, patterns, and vulnerabilities. The process guides an analyst through three security analysis phases with an improved weakness (CWE-1000) and entirely new architecture pattern databases. However, the method is still subject to refinement and tuning.
	
	\section{Method}
	\label{sec:method}
	
	\begin{figure}
		\centering
		% insert flowchart-figure here to show
		\begin{tikzpicture}[%
			>=triangle 60,              % Nice arrows; your taste may be different
			start chain=going below,    % General flow is top-to-bottom
			node distance=6mm and 60mm, % Global setup of box spacing
			every join/.style={norm},   % Default linetype for connecting boxes
			% Parameters for scaling tikz
			%thick,
			scale=0.7,
			every node/.style={transform shape}
			]
			% ------------------------------------------------- 
			% A few box styles 
			% <on chain> *and* <on grid> reduce the need for manual relative
			% positioning of nodes
			\tikzset{
				base/.style={draw, on chain, on grid, align=center, minimum height=4ex},
				proc/.style={base, rectangle, text width=8em},
				test/.style={base, diamond, aspect=2, text width=5em},
				term/.style={proc, rounded corners},
				% coord node style is used for placing corners of connecting lines
				coord/.style={coordinate, on chain, on grid, node distance=6mm and 25mm},
				% nmark node style is used for coordinate debugging marks
				nmark/.style={draw, cyan, circle, font={\sffamily\bfseries}},
				% -------------------------------------------------
				% Connector line styles for different parts of the diagram
				norm/.style={->, draw, lcnorm},
				free/.style={->, draw, lcfree},
				cong/.style={->, draw, lccong},
				it/.style={font={\small\itshape}}
			}
			% -------------------------------------------------
			% Start by placing the nodes
			\node [proc, densely dotted, it] (p0) {Smart-Lighting System};
			% Use join to connect a node to the previous one 
			\node [term, join] (p1) {Start of analysis};
			\node [proc] 	   (p2) {Extract specifications};
			\node [proc, join] 		{Role assignment};
			\node [proc, join]      {Pick related domain references};
			\node [proc, join] (l1)	{Map layers to components};
			\node [proc, join] (p3)	{Layer cross-mapping};
			\node [test, join] (t1) {Mapping complete?};
			\node [proc] 	   (p4) {Create multi-domain taxonomy};
			
			\node [proc, densely dotted, it, below right=of p0, xshift=-15mm] (e1) {Domain layered architecture articles};
			
			% Next column
			\node [proc, right=of p2] (p5) {Attack feasibility validation};
			\node [proc, join] (p6) {Differential description};
			%			\node [test, join] (t2) {All layers covered?};
			\node [proc, join] (p7) {Application-specific attack tree};
			\node [proc, join] (p8) {Derive threats and vulnerablities};
			\node [term, join] (p10) {End of analysis};
			% -------------------------------------------------
			\node [coord, left=of p4] (c0)  {}; 
			\node [coord, right=of p2] (c1)  {};
			\node [coord, below=of c1] (c11)  {}; \cmark[1]{1}
			%			\node [coord, right=of p5] (c2)  {}; 
			%			\node [coord, above right=2mm of c2] (c21)  {}; \cmark[1]{2}
			% -------------------------------------------------
			% A couple of boxes have annotations
			\node [above=0mm of p2, it] {(Identification)};
			\node [above=0mm of p5, it] {(Classification)};
			% -------------------------------------------------
			% All the other connections come out of tests and need annotating
			% First, the straight north-south connections. In each case, we first
			% draw a path with a (consistently positioned) annotation node, then
			% we draw the arrow itself.
			\path (t1.south) to node [near start, xshift=1em] {$y$} (p4); 
			\draw [->,lcnorm] (t1.south) -- (p4);
			%			\path (t2.south) to node [near start, xshift=1em] {$y$} (p7); 
			%			\draw [->,lcnorm] (t2.south) -- (p7);
			
			\draw [->,lcnorm] (p4.east) -- ++(15mm,0mm) |- (p5);
			\path (t1.east) -| node [very near start, yshift=-1em, xshift=2mm] {$n$} (c1); 
			\draw [->,lcfree] (t1.east) -| (c1)  |- (p2);
			%			\path (t2.east) -| node [very near start, yshift=-1em, xshift=2mm] {$n$} (c2); 
			%			\draw [->,lcfree] (t2.east) -| (c2) -- (p5); 
			% -------------------------------------------------
			\draw [-,lcnorm, gray, densely dotted] (p1.south) -- (p2.north);
			\draw [->,lcnorm] (e1.west) -| ($(e1.west)+(-3mm,0mm)$) |-  (p1.east);
			
			% -------------------------------------------------
			% Loop arrows
			\draw [->, green!50!black] ($(l1.south east) +(2mm,1mm)$) arc (20:360:3mm);
			\draw [->, green!50!black] ($(p3.south east) +(2mm,1mm)$) arc (20:360:3mm);
			\draw [->, green!50!black] ($(p4.south east) +(2mm,1mm)$) arc (20:360:3mm);
			\draw [->, green!50!black] ($(p5.south east) +(2mm,1mm)$) arc (20:360:3mm);
			\draw [->, green!50!black] ($(p6.south east) +(2mm,1mm)$) arc (20:360:3mm);

			% -------------------------------------------------
			% Highlighting boxes 
			\draw[orange,thick,dotted] ($(p2.north west)+(-5mm,5mm)$) rectangle ($(p4.south east)+(5mm,-3mm)$);
			\draw[brown,thick,dotted]  ($(p5.north west)+(-5mm,5mm)$) rectangle ($(p8.south east)+(5mm,-3mm)$);
			
			% -------------------------------------------------
			% 
			\draw [->, green!50!black] ($(p4.east) +(30mm,2mm)$) arc (20:360:3mm);
			\node [black,right=30mm of p4, it] {\shortstack{denotes iterative tasks done,\\once for each reference model}};
		\end{tikzpicture}
		\caption{Design cycle steps carried out for a Smart-* system vulnerability identification. {\small }}
		\label{fig:flowchart}
	\end{figure}
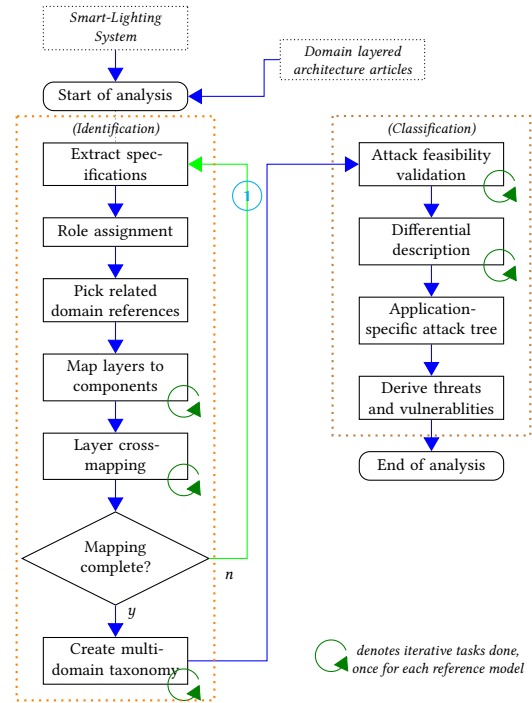
	
	For our layer-based technique, we processed the target \textit{Smart-Lighting} architecture (SLA) in all its components, Figure~\ref{fig:flowchart}. Inspired by design science~\cite{Hevner2007}, our technique combines results and experiences from a knowledge base of related domains. An iterative design process (design cycle) integrates them to create artifacts assessed in-field (relevance cycle). Finally, the present technique and resulting artifacts will grow the knowledge base (rigor cycle). The focus of this paper's design process is described in the following consists of two steps: identification and classification.  
	
	\textit{Identification.} As with any modern system-of-systems, an SLA contains multiple heterogeneous systems, each with its domain-specific constraints. Hence, we first analyze the system's composition by gathering specifications from technical data sheets and reconstructing its architecture diagram. In particular, each component gets assigned one or more architectural roles. For example, we will see that the LoRaWan end-node controllers take up two roles, (B) in Figure~\ref{fig:aLayout}. They act as a communication gateway (networking) and perform decentralized supervision of the connected lighting-bus devices (control). 
	
	Domain-specific security aspects further characterize a component. For example, physical tampering characterizes a light device. Based on previous work~\cite{Hofer2018}, we select domain-specific layered architectural reference models ($RM_i$) from representative research papers. Each model holds a different architectural focus (e.g., control flow) or domain (e.g., IoT) and carries related information on possible layer-level attacks and vulnerabilities. 
	
	\begin{table*}[tb]
		\centering
		\caption{Example LoRaWan end-node layer role and definition differences for reference models}
		\small
		\begin{tabular}{l p{0.23\linewidth} p{0.23\linewidth} p{0.30\linewidth}}
			\toprule
			Role & {$RM_A$}~\cite{AshibaniMahmoud2017}  & {$RM_H$}~\cite{Hanetal2014} & {$RM_L$}~\cite{Linetal2017-2}\\ 
			\midrule
			Control & Application: [..] process the received information from the data transmission level and issue commands to be executed by the physical units, sensors and actuators.
			& Supervisory Control: By aggregating the measurement data from multiple points in the network, the supervisory sub-control level creates system-level feedback control loops, which make system-level control decisions.
			& Application: [..] receives the data transmitted from network layer and uses the data to provide required services or operations. For instance, the application layer can provide the storage service [..] or provide the analysis service to [..] predicting the future state of physical devices.\\
			Communication & Transmission: is responsible for interchanging and processing data between the perception and the application. [..] are achieved using local area networks, communication networks, the Internet or other existing networks [..].
			& Network: [..] takes charge of networking sensors and actuators as well as bridging the sensor/actuator layer and the higher control layer with a variety of communication devices and protocols.
			& Network: [..] used to receive the processed information provided by perception layer and determine the routes to transmit the data and information to the IoT hub, devices, and applications via integrated networks.\\
			\bottomrule
		\end{tabular}
		\label{tab:mapdef}
	\end{table*}
	Based on component roles, we draw a map $M_i$ between the components and the layers of a model $\Lambda (RM_i)$. For example, the sensor and actuator layer contains a light device. Starting bottom-up, we iterate through the architecture components and survey each $RM$ for matching role descriptions. We ensure that: a) every component fits into at least one layer of a reference model, and b) for each layer, there exists a component that maps into it. The former ensures that each component can be described in each reference model and enriched with the information of a layer's attacks and vulnerabilities. The latter secures that all attacks described in each reference model find a target in our system. As an example, Table~\ref {tab:mapdef} shows the roles of a LoRaWan lighting controller and their assigned layer mapping among three reference models. For $RM_A$'s generic CPS and IoT-oriented and $RM_L$'s service-oriented model, the "Control" role maps best to their Application layer. In contrast, in the control-oriented $RM_H$ model, this supervising component best fits the Supervisory Control sub-layer.  While layer descriptions are similar, the focus diverges slightly between models. Also, the attack definitions for the mapped layers reflect such proximity. For Example, the definitions for \emph{Malicious Code} ($RM_H$) and \emph{Malicious virus/worm} ($RM_L$) refer to the same type of attack. However, they diverge due to focus, i.e., performance vs. data-centric, emphasizing the importance of creating a unified taxonomy. 
	
	Each component now equips a role and projected attacks for each mapped reference model layer. Consequently, we can link layers from the different reference models through their common component mapping. This consolidation phase constructs thus cross-mappings $CM_{ij}$ among the reference model layers $\Lambda(RM_i)$ and $\Lambda(RM_j)$ based on component allocation.	However, as one model's layer definition may encircle only a subset of another model's layer, cross-mappings are unidirectional and may not hold in reverse. It is typically the case for more abstract models that cross-map to layer-rich models. Therefore, we suggest starting with the former as it allows for easier and better associations.

	%TODO dummy table LoRa Controller with definition integration example?
	%NOTE: table showing functions of RM-CM-LRM?
	
	In the final identification step, we use the cross-mapping $CM_{ij}$ to create a differential attack and threat table. This table contains relevant definitions to verify a Smart-Lighting system for attacks, threats, and vulnerabilities. First, we create an initial base table that records layer relationships and enriches each entry with its original research's attack taxonomy. We further append each attack's origin and original meaning for a more accurate understanding by following citations. Next, starting again from the least detailed model, e.g., $RM_A$, we review attack definitions of layers connected by the cross-mapping, $CM_{ij}$. We remove duplicates, integrate definitions, or highlight differences or ambiguities. If an undefined term appears, we define it according to its context with the help of other domain-related and reference sources. Once complete, we group the remaining entries by layers and threads to build the differential, application-specific taxonomy table. Consequently, the consolidation outputs are a layer mapping to the architecture and between models and a taxonomy table that includes the analyzed multi-domain perspective.
	
	\textit{Classification.} With the preceding table, we perform a differential weakness discovery for the SLA. We evaluate each attack's definition and assess if, in the Smart-Lighting domain, the proposed attacks remain possible or sensible. Starting bottom-up in the architecture, we pick a network or its next component and verify each attack in the differential taxonomy table for every assigned layer in the model. The process repeats until it analyzes all reference model layers and SLA components. We summarize and discuss the results of the attack analysis in a differential description that presents newly found attacks while honoring a previous model's results. 
	
	To highlight differences and commonalities to CPS of a Smart-Lighting system, we create a domain-specific attack tree based on the generic tree created by Alguliyev \textit{et al.}~\cite{Alguliyevetal2018}. Then, using their attacks-threats functional CPS model, we reuse or define further attacks and threats in the taxonomy derived from our study. The reviewing of vulnerability definitions of the reference articles and the resulting attack tree will then serve as input for a final assessment of the possible vulnerabilities in a Smart-Lighting system.
	
	\section{Identification}
	\label{sec:identification}
	
	\subsection{Smart-Lighting architecture under study}
	
	\begin{figure}[tb]
		\centering
		\includegraphics[width=\linewidth, height=0.45\textwidth,keepaspectratio]{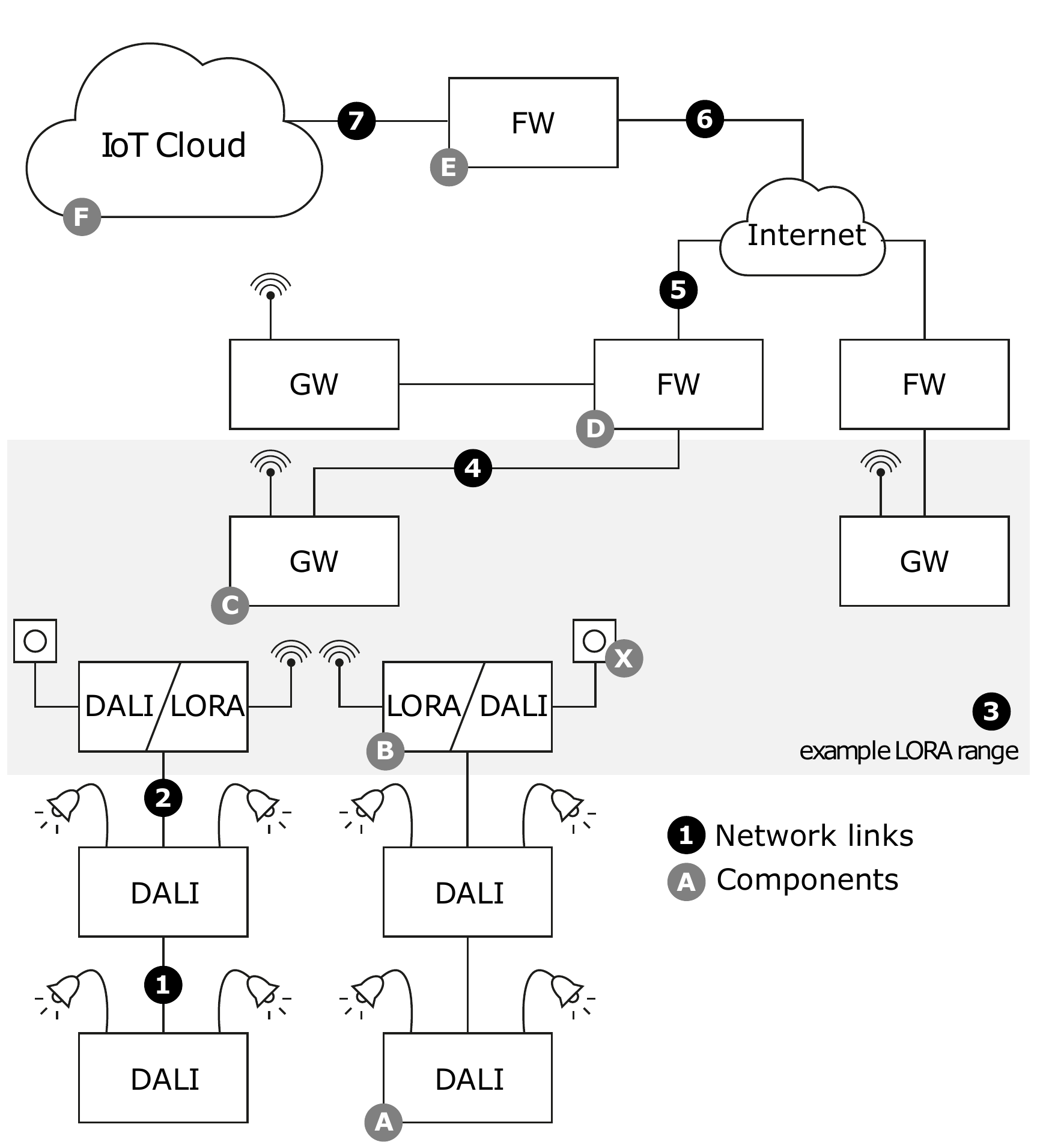}
		\vspace{0px}
		\bigskip
		
		\includegraphics[width=0.45\textwidth, height=0.5\textwidth,keepaspectratio]{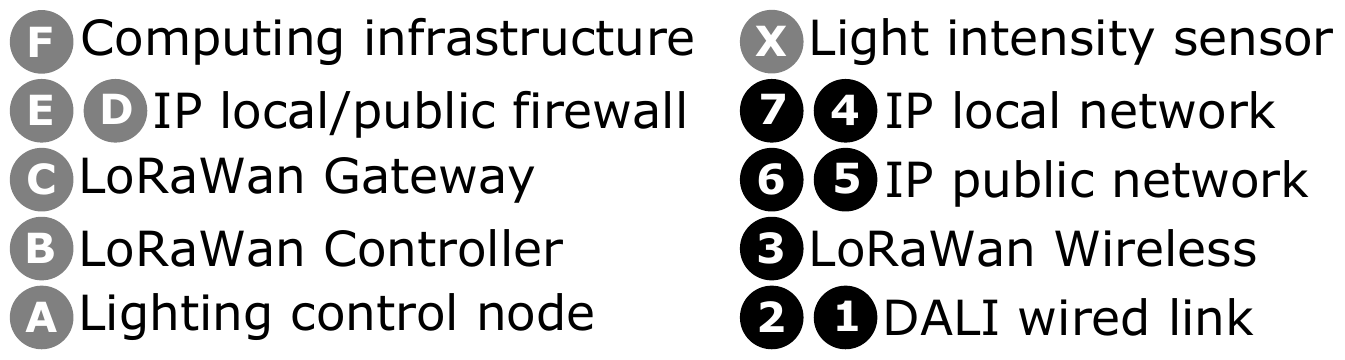}
		\caption{Smart-Lighting case study architecture}
		\label{fig:aLayout}
		\vspace{-10px}
	\end{figure}
	To explain our method, we use the case study architecture of a Smart-Lighting installation. The \textit{Smart-City} pilot project runs in the city of Merano, Italy. It covers an area of $26 km^2$ and more than 6.700 distributed light posts. Figure~\ref{fig:aLayout} illustrates the result of the identification step. The figure shows a simplified version for the demonstrative purpose of the installed system's architecture containing all the elements needed to create smart, remotely controlled lighting infrastructure. Three different networking technologies convey the control and status information between the end-nodes and the computing cloud: light devices (1-2), wireless network (3), and a traditional IP-based network (4-7). Each of the three networks varies in function and timing constraints, acting thus as a leveled control model~\cite{Hallmansetal2015}.
	
	\textit{Dali end-nodes.} Digital Access Light Interface (DALI), a master-slave two-wire message-based bus for lighting and illumination systems, interconnects the light devices~\cite{Bellido-Outeirinoetal2012}. Its self-clocked differential encoding runs the data on a low data rate of 1200baud in half-duplex, and when externally powered, for multiple hundred meters and resilient to interference~\cite{Hein2001, Bellido-Outeirinoetal2012}. The DALI end device controllers (A), also called ballast controllers, execute simple application-specific programs and require only small micro-controllers~\cite{Contenti2002}. In 2017, the Digital Illumination Interface Alliance (DiiA) released a revised version of the standard. DALI 2 standardizes timing requirements and signal slopes, increasing interoperability~\cite{DiiA2018}. It also adds multi-master operation or multiple logical units per bus device while maintaining backward compatibility with DALI 1.
	
	\textit{The LoRaWan network.} The wireless star-of-stars network is designed on a LoRaWan (Long Range Wide Area Network) master-slave protocol that runs on top of a Semtech LoRa wireless transmitter~\cite{Lora2017}. The transmitter operates in Industrial-Scientific-Medical (ISM) band with either 250 half-duplex channels of 5.5 kbps and one at 11 kbps in chirp spread spectrum (CSS), or one channel of 50 kbps in frequency shift key (FSK) modulation (Europe channels). Its transmission robustness outperforms traditional systems, enabling servicing thousands of devices and reducing the need for a mesh network~\cite{Augustinetal2016}. The LoRaWan network associates nodes (B) through gateways (C) to network and application servers (F). A LoRaWan end-node (B) can have different modes: event-driven sensors, beacon scheduled actuators (usually both battery-powered), or always online. It stores two AES128 keys, securing the communication to the \emph{network} and \emph{application} server. The installed gateways (C) serve as bi-directional relays and mount multichannel-multi-modem units for simultaneous reception on different frequencies and data rates without any end-node association or handover. A network server manages the distribution of data flow between an application and nodes. It reconfigures a gateway's multi-modem channels and data rate according to needs and environmental conditions. Such an adaptation targets the shortest air time (Adaptive data rate) and the best channel diversity (Channel maps) while increasing overall transmission efficiency and total throughput~\cite{Lora2017}. As the entry and exit-point of data flow are not binding, LoRaWan supports redundancy by default. However, the network setup makes direct end-node communication impossible~\cite{Augustinetal2016}. The used LoRaWan end-device mounts a LoRaWan/DALI master controller for routing and the timed control of connected DALI devices, and Bluetooth LE hardware for the initial configuration setup~\cite{UL20x0}. It offers over-the-air (OTA) firmware update and OTA device activation and features digital and analog inputs and outputs to attach optional sensor-actuator hardware. The hardware of the used LoRaWan gateway mounts an ARM Cortex-A\texttrademark processor running a Linux kernel. It allows user program deployment and features a backup uplink over 3G.
	
	\textit{The IP infrastructure.} The IP-based infrastructure is configured as in a traditional IT system. Local networks use IPv4 or IPv6 connectivity through Internet (5-6) and unite firewalls with gateways (4) and computation cloud (7). Within and between networks, standard protocols (IPSec/HTTPS) secure connections. The firewalls (D-E) perform routing and protection tasks, providing traditional intrusion detection algorithms. The cloud environment (F) stores and analyzes data. The data from the on-site gateways enters the cloud through a software firewall, which forwards it to the ``Loriot IoT'' network server running as an IaaS instance. The latter forwards the message payload to a PaaS application server operating an ``Azure IoT'' service running a set of custom-developed ``Azure Functions'' and micro-services. These gather and store the acquired information in a ``Kosmos DB'' No-SQL database and take control measures accordingly. The virtual LAN and firewall (E) configuration allow internal data flow governance and additional fine-grained protection mechanisms.
	
	\subsection{Layer mapping}

	To cover all aspects of our SLA, we use four reference models, $RM_i$, originating from varying domains: one model to cover generic aspects of CPS's information security, one to highlight and stress the importance of information and control flow in CPS, and two to extend aspects peculiar to IoT, Big Data, and service orientation. The SLA maps then to the reference models. Reference models $RM_i$, their mappings $M_i$ and cross-mappings $CM_{ij}$ are further detailed in the rest of this section.
	
\begin{figure}[t]
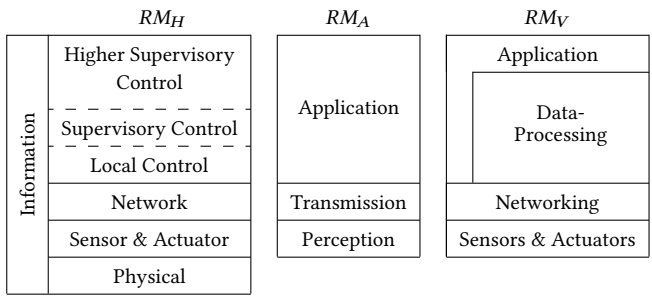

		\centering
		\small
		\renewcommand{\arraystretch}{1.4}
		\begin{tabular}{| c | c | c | c | c | c  c |}
			\multicolumn{1}{ c }{} & \multicolumn{2}{ c }{$RM_H$}& \multicolumn{1}{ c }{$RM_A$} & \multicolumn{1}{ c } {}&
			\multicolumn{2}{ c }{$RM_V$}\\ 
			
			% Top lines
			\cline{1-2}\cline{4-4}\cline{6-7}
			
			\parbox[t]{2mm}{\multirow{7}{*}{\rotatebox[origin=c]{90}{Information}}} & \multirow{2}{*}{\shortstack{Higher Supervisory\\Control}} & & \multirow{4}{*}{Application} & & \multicolumn{2}{|c|}{Application} \\
			\cline{7-7}
			
			& & & & & & \multicolumn{1}{| c |} {\multirow{3}{*}{\shortstack{Data-\\Processing}}}\\ 
			\cdashline{2-2}
			
			& Supervisory Control & & & & &\multicolumn{1}{| c |}{ }\\ % Han LC = Dali, Supervisor = LoraControl, HC = IoT cloud
			\cdashline{2-2}
			
			& Local Control  & & & & &\multicolumn{1}{| c |}{} \\
			\cline{2-2}\cline{4-4}\cline{6-7}
			
			& Network & &  Transmission & & \multicolumn{2}{| c |}{Networking}\\
			\cline{2-2}\cline{4-4}\cline{6-7}
			
			& Sensor \& Actuator & & Perception & & \multicolumn{2}{| c |}{Sensors \& Actuators} \\
			\cline{2-2}\cline{4-4}\cline{6-7}
			
			& Physical & \multicolumn{1}{ c }{} & \multicolumn{4}{ c } {} \\
			\cline{1-2}
			
		\end{tabular}
		\caption{Comparison of layer allocations for the SLA of Figure~\ref{fig:aLayout} in relation to $RM_A$, the least detailed model}
		\label{fig:layers}		
	\end{figure}

	$RM_H$, (Han \textit{et al.}~\cite{Hanetal2014}): the architecture of systems arranges in a 4 plus 1 layer model, Figure~\ref{fig:layers} center. Its layer stack contains the Physical, Sensor and Actuator, Network, and Control layer. The latter divides into three control-oriented sub-layers: Local-distributed control action layer, Supervisory sub-control level, and Supervisory higher control level. This division highlights hierarchical separation and enables distributed independent control. The plus one (Information) layer interfaces transversely, acting on all four stacked layers. It represents the information flow sinking to the top and sourcing from among all layers in the architecture or vice versa, supporting the notion of shared information for distributed control.
	
	$RM_A$, (Ashibani and Mahmoud~\cite{AshibaniMahmoud2017}): uses a three-layer approach defined as the Perception, Transmission, and Application layer. Its architectural distribution is similar to $RM_H$ in that both propose a centrally layered stack with similar features. While the authors acknowledge that three layers are not enough to abstract all CPS functionality, the model suffices to capture the functional core. Without a Physical and an Information layer, $RM_A $ proposes a more generalized view that allocates all control and computation on the top layer. 
	
	$RM_V$ (Varga et al.~\cite{Vargaetal2017}): focus on IoT and distributed data acquisition. It draws on the previous three-layer model but adds a Data processing layer to take care of the vast data mole entering the IoT hub. This addition suggests a strong focus on data processing and process automation analytics. 
	
	$RM_L$, (Lin et al.~\cite{Linetal2017-2}): details the aspect of service orientation in an IoT-based layered architecture. Similar to $RM_V$, they extend the three-layer model with an additional Service-oriented layer between Network and Application. The layer orchestrates and manages the available services to translate, process, and store incoming and outgoing data. As this role is passive, it suffers from adjacent layers' vulnerabilities, making it transparent. As we will see during mapping, $ RM_L $ ends up virtually equivalent to $ RM_A $.
	
	\begin{table}[tb]
		\caption{Identified architecture roles and description}
		\begin{tabular}{l p{0.6\linewidth}}
			\toprule
			Role & Description\\
			\midrule
			Store \& Process & Elaboration and analysis of information \\
			Control	& Feedback management informed via sensing and actuating\\
			Communication &	Data exchange and bridging between locations, systems and technologies.\\
			Sensing & Capturing and quantifying of the physical world \\
			Actuating & (Inter)action on the physical world.\\
			\bottomrule
		\end{tabular}
		\label{tab:roles}
	\end{table}
	Next, we compose layer mapping ($M_i$) for each model and component in Figure~\ref{fig:aLayout} based on the identified roles in Table~\ref{tab:roles}. An integrated streetlight (A) senses the lamp current and actuates lamp illumination levels. A generic sensor, such as a light intensity sensor (X), captures light intensity. We map both thus to $RM_H$'s Sensor and Actuator layer. $RM_V$'s Sensors and Actuators layer matches the same description, while the Perception layer offers the best match for $ RM_A $ and $ RM_L $. The light device further uses lamp data and control to monitor and govern light intensity and system health. In $ RM_H $, the Local(-distributed) Control sub-layer manages the given sensory information locally, acting as a local control entity. All other $RM$ refer to a single Application layer for this purpose. Figure~\ref{fig:aLayout} shows these devices connected to peer nodes and a master controller (B) through wired couplings (1)-(2). The latter firstly functions as a network bridge between DALI and LoRaWan. It forwards the information over wireless connections (3) and IP networks (4-5-6-7) through firewalls (D-E) and gateways (C) to the IoT Cloud (F). The Network layer of $ RM_H $ and $ RM_L $ best describes these devices' and links' connectivity roles. It is responsible for distributing and interconnecting devices, sensors, actuators, and services the Control layer. Similar descriptions for the Transmission layer in $ RM_A $ and the Networking layer in $ RM_V $ fall into place. Secondly, LoRaWan end-node controllers (B) perform minor decentralized supervision, switching, and timing operations of the connected light devices. This description maps to the Supervisory Control sub-layer of $RM_H$ to which all local controllers subside. The nodes report back to a higher instance, a business process located at the IoT cloud. This process is in charge of management and control of the system's overall operation, i.e., the city, and relates to the Supervisory Higher Control of $ RM_H $. All other $RM$ refer to both mentioned control sub-layers to the single Application layer. $ RM_V $ and $ RM_L $, however, have different role associations for the IoT cloud. $ M_V $ includes an assignment of the Data processing layer in charge of information pre-processing. At the same time, $ M_L $ foresees a Service-oriented architecture to manage service interaction and processes. All the above components handle or contain information of some sort. $ RM_H $'s Information layer applies thus on all components and roles. Further, the physical world is specified only in $RM_H$ mapped, thus, to the physical layer. Table~\ref{tab:devicelayers} summarizes the device-layer assignments of this paragraph.
	
	\begin{table*}[t]
		\centering
		\caption{Device layer allocations for the SLA of Figure~\ref{fig:aLayout}}
		\small
		\begin{tabular}{ l c c c c c c c c c }
			\toprule
			Device	& & $RM_A$ & & \multicolumn{2}{ c }{$RM_H$} & & $RM_V$ & & $RM_L$ \\
			\midrule
			
			\multirow{3}{*}{IoT Cloud} & \multirow{3}{*}{(F)} & Application & & \parbox[t]{2mm}{\multirow{14}{*}{\rotatebox[origin=c]{90}{Information}}} & \shortstack{Higher supervisory\\ Control} & &  Application & & Application \\
			\cline{3-3}\cline{6-6}\cline{8-8}\cline{10-10}
			
			& & -- & & & -- & & Data processing & & -- \\
			\cline{3-3}\cline{6-6}\cline{8-8}\cline{10-10}
			& & -- & & & -- & & -- & & Service-oriented \\
			\cline{3-3}\cline{6-6}\cline{8-8}\cline{10-10}
			& & \multirow{4}{*}{Transmission} & & & \multirow{4}{*}{Network} & & \multirow{4}{*}{Networking} & & \multirow{4}{*}{Network} \\
			\cline{1-1}
			
			Firewall & (D-E) & & & & & & & & \\
			\cline{1-1}
			
			Gateway & (C) & & & & & & & & \\
			\cline{1-1}
			
			\multirow{2}{*}{Controller} & \multirow{2}{*}{(B)} & & & & & & & & \\
			
			\cline{3-3}\cline{6-6}\cline{8-8}\cline{10-10}
			& & \multirow{2}{*}{Application} & & & Supervisory Control & & \multirow{2}{*}{Application} & & \multirow{2}{*}{Application} \\
			\cline{1-1}\cline{6-6}
			
			\multirow{2}{*}{Lighting node} & \multirow{2}{*}{(A)} &  & & & Local Control & &  & &  \\
			\cline{3-3}\cline{6-6}\cline{8-8}\cline{10-10}
			& & \multirow{2}{*}{Perception} & & & \multirow{2}{*}{\shortstack{Sensor and\\Actuator}} & & \multirow{2}{*}{\shortstack{Sensors and\\Actuators}} & & \multirow{2}{*}{Perception} \\
			\cline{1-1}

			Light Sensor & (X) & & & & & & & & \\
			\cline{1-1}\cline{3-3}\cline{6-6}\cline{8-8}\cline{10-10}

			\cline{3-3}\cline{6-6}\cline{8-8}\cline{10-10}
			
			\cline{3-3}\cline{6-6}\cline{8-8}\cline{10-10}
			\cline{1-1}
			
			Physical world & & -- & & & Physical & & --  & & -- \\
			
			%			\multirow{2}{*}{--}  & & \multirow{2}{*}{--} \\
			%			\cline{1-1}
			%			
			%			Physical world &  & & & & & & & & \\
			
			\bottomrule			
		\end{tabular}
		\label{tab:devicelayers}		
	\end{table*}
	
	$CM_{AH}:$ The perception layer maps to the Sensor and Actuator layer from Han \textit{et al.}, the Transmission to the Network layer, and the Control to the Application layer. However, $RM_A$ has no reference neither for the physical nor for the information layer. While the latter might blend into the existing three layers of $ RM_A $, no notion of physical components other than sensors or actuators is present in $ RM_A $, making this a partial extending cross-mapping.
	
	$CM_{AV}, CM_{AL}:$ The two IoT models show only minor mapping differences to the three-layer model of $RM_A$. Both map almost directly with minor differences in naming for Transmission/Networking and Perception/Sensor and Actuator. The fourth layer in both proposals shares some functionality with their lower layer. However, it communicates to the upper layer, partially parallel to $ RM_A $'s application layer. Their function distribution on the example architecture emerges thus almost identical and transparent. These models can thus extend $RM_A$s notions of attack for application layers with details based on data processing.
	
	\begin{table*}
		\centering
		\caption{Excerpt from the Perception layer of the multi-domain taxonomy table, with notes of $RM$ origin}
		\begin{tabular}{p{0.15\linewidth} p{0.2\linewidth} p{0.4\linewidth} p{0.11\linewidth}}
			\toprule
			Threat & Attack & Note & Also seen \\
			\midrule
			Resource exhaustion	& Path based DoS (A) &	Sending messages along the routing path, flooding & \\	
			& Flooding (A) &	Flooding a service or resource with requests, ex join &	Flooding (V) \\
			& (D)DoS attacks (A) &	Attempt to make network or service unavailable & \\	
			& Replay attack (A)	& Replay authentication to deplete authentication authority resources; also delayed, to external entities or to other node to gain access or trust (H) & Replay (H), Replay (V), Replay (L)\\
			\bottomrule
		\end{tabular}	
		\label{tab:taextract}
	\end{table*}
	
	Finally, we build a common taxonomy starting with the most abstract model $RM_A$. First, we compare definitions with the more detailing classification of $RM_H$ and the definition extensions for service and data-centric architectures of $RM_V$ and $RM_L$. Then, we align definitions with $RM_A$'s layers in a separate spreadsheet, mark inconsistencies, additions, duplicates in color, and finally filter and merge them. It is worth noting that we encountered and marked multiple definitions during this step with untraceable origin. The grouped remaining attacks by layer and threat, where we use $RM_A$ as layer base, result like the excerpt in Table~\ref{tab:taextract}. The table lists threats for each layer, filtered attacks, duplicates, and the definitions, possibly integrated with the duplicate's details. For example, the last row lists three attack duplicates in other $RM$s. It integrates the $RM_A$ definition of ``Replay attack`` with detail from $RM_H$.
	
	Figure~\ref{fig:layers} displays the cross-comparison result of the consolidation phase. All tables containing attack and threats descriptions of the different stages of this work are available as download\footnote{~Industry 4.0 - Smart-Lighting Taxonomy table \url{https://bit.ly/3nHaxjN}}. 
	
	\section{Classification} 
	\label{sec:classification}
	
	In this section, we iterate through the SLA in Figure~\ref{fig:aLayout} and verify if the attacks remain possible or sensible. As the capabilities of the three network technologies differ significantly, we analyze each network separately by component. Then, through their component's layer allocations (Table~\ref{tab:devicelayers}), we verify the feasibility of a layer's attack using the unified differential taxonomy. Finally, we determine threats and connected vulnerabilities, concluding with suggestions for countermeasures.
	
	\subsection{Attack validation}
	\label{sec:attacks}
	\subsubsection{DALI Network}	
	\label{sec:dali}
	The DALI network consists of streetlights (A) interacting with the physical world, connected through a two-wire bus (1-2) to a LoRaWan end-node (B) that acts as network master, Figure~\ref{fig:aLayout}.
	
	\textit{$RM_A$~\cite{AshibaniMahmoud2017}:} At the Perception layer, we mainly see two types of node attacks: attacks that \textit{physically} act on the node and attacks that \textit{virtually} interact with the node. The former requires some form of physical activity on the node where an attacker can get, alter, or make information inaccessible through node capture, tampering, or destruction. The latter type interferes with the node's function by intervening in sensor measurement or corrupting data integrity. Physical attacks may cause information disclosure by, e.g., replacing the node with a duplicate, stealing its data, replicating its functions, and attacking information integrity through false information. Such attacks may cause system malfunction, e.g., darken specific city areas, as lampposts are publicly accessible. For virtual attacks on systems using the new DALI 2 standard, a captured or inserted false node could act as master and take over other nodes (i.e., spoofing). It may actively poll or even change another node's data and configuration, directly controlling, corrupting, and desynchronizing a network. DoS attacks, such as flooding, can take out a node and make its services unavailable. Finally, electromagnetic interference attacks influence sensory measurements and actuation control, e.g., through action on the system's resonance frequency, corrupting measured values, or feedback loops. 
	
	Misleading attacks and buffer overflow occur at the application layer in (A) and (B). The former attacks attempt to make status or value readings unreachable (Denial of Service -- DoS), forge commands, or intercept and manipulate loops through altered information (Man in the Middle -- MitM). The latter inject malicious code. Such attacks to be successful require specialized knowledge of the attacked micro-controllers~\cite{Contenti2002, UL20x0}. On this layer, all attacks interact virtually with the node. Thus, an attacker with network access can systematically trial all reachable nodes.
	
	Virtual attacks gain even more visibility on the Transmission layer of a controller (B). Namely, physical access to the two-wire DALI bus allows an attacker to perform DoS or selective collision attacks (including the mere cutting of wires), muting targeted nodes, and disrupting or desynchronizing control loops. Flooding, or attacker-initiated off-the-schedule polls, can quickly exhaust the network's limited relay capacity. While the simple bus intrinsically avoids routing-based vulnerabilities such as MitM or selective forwarding, its standard lacks authentication and encryption. That enables eavesdropping or, on the new DALI 2 standard, data tampering and forging control messages. Finally, it is worth mentioning that an attacker can orchestrate most of the listed attacks remotely through, e.g., a captured LoRaWan gateway node (B)
	
	\textit{$RM_H$~\cite{Hanetal2014}:} The model redefines the desynchronization attack (called Control Forgery in $ RM_A $) for the Control layer. The new definition calls it specifically designed to damage a system, e.g., delayed instrument readings that dis-align physical and cyber worlds. In $ RM_A $, it only causes generic system misbehavior. For the Network layer, $ RM_H $ suggests the spoofing attack may also aid in transmitting false error messages. These messages suggest fictitious lamp failures to the supervisory control, disabling the lamp. On the Physical layer, attacks to external system components are considered. An attacker may intervene on DALI infrastructure and hinder its operation, e.g., cover a lamp. Finally, the Information layer highlights privacy issues that might arise through the information extracted from the transmitted data. For example, the presence/passage of persons in motion-activated areas hints at vacancy. It may cause burglary of adjacent housing units. 
	
	\textit{$RM_V$~\cite{Vargaetal2017}:} 
	The Application layer of this model adds configuration tampering attacks for both nodes. Due to resource constraints or programming errors, the embedded code on ballasts and LoRaWan end-nodes might not verify control parameters for limits and constraints. Such attacks set invalid operating values, e.g., default illumination values to zero, disabling illumination, and threatening safety. 
	
	\textit{$RM_L$~\cite{Linetal2017-2}:} This model considers unauthorized users' implications on the network level. Like configuration attacks, unprotected DALI allows an attacker to alter device settings with comparable results (Perception layer). Furthermore, the model identifies malicious code injection attacks as a source of access for multiple system levels. Thus, the node would act as a vehicle for diffusion on all levels. However, as for the attack in $ RM_A $, resource constraints and specificity make it hard to predict their success.
	
	\subsubsection{LoRaWan Network}
	\label{sec:lora}
	A typical SLA combines multiple gateways (C), LoRaWan end-nodes (B), and at least one Network server (F) through LoRa (3) to create a city-wide LoRaWan network, Figure~\ref{fig:aLayout}. In addition, these LoRaWan end-nodes may feature sensors and actuators (X) which interact with the physical world for further illumination control or monitoring.
	
	\textit{$RM_A$~\cite{AshibaniMahmoud2017}:}	On the Transmission layer, the LoRaWan network exposes multiple availability-related attacks as adversaries can directly access the ether. For example, despite the robust multi-channel multi-modem gateway configuration, typical {DoS} attacks are feasible through multi-channel frequency jamming, intentional collision, or flooding. Notably, a random message flooding attack targeting those gateways might disrupt a network section. However, the latter solely react to preambles and, by default, cannot handle more than ten packets at a time~\cite{Semtech2017-2}. Furthermore, a targeted DoS attack will cause collisions and force re-transmissions, ultimately exhausting a battery-powered node's energy.	
	
	Suppose these messages are further ``replays'' of join requests (replay attack). In that case, forwards to Join or Network servers add computation burden and eventually exhaust available resources. Collision attacks have a similar overpowering effect. Unverified transmission practice on the medium and its protocol similarities to ALOHA impact severely on successful message reception, i.e., channel exhaustion at 60\% load and only 18\% of total capacity~\cite{ElFehrietal2018, Augustinetal2016}. A jamming attack is harder to perform and requires at least three parallel transmissions on the default LoRaWan frequencies close to a device~\cite{Lora2017}. Namely, an adequately configured device will see jamming as radio interference and re-transmit on a different channel. Related attacks, such as a resonance attack, will also identify as interference and cause the same response~\cite{Augustinetal2016}. However, if unconfirmed messaging is used, the data loss may go unnoticed and not trigger any change. Listen-in and analyzing this high number of re-transmitted packages enables side-channel and time analysis attacks to deduce session-key composition. However, the used two-layered encryption limits attack effectiveness. Other typical attacks for the Transmission layer, such as MitM, Sybil, and eavesdropping, remain ineffective until successful capture of the key. Despite missing keys, traffic analysis helps conclude origin, network configuration, and message function. Alternatively, an attacker can attempt node capture and tamper with its memory. The node hosts the necessary keys to send manipulated messages, opting to disrupt the network using valid credentials.
	
	A node (B) is again subject to misleading and buffer overflow attacks at the Application layer. However, while session-keys-protected channels harden a manipulation resorting to MitM, command forgery and interception,  attempts to make status or value readings unreachable (DoS) stay valid. Thus, even though transmission requires master capabilities and session keys, the same risks for code injection as for the DALI network apply. Again, an attacker with network access can systematically trial all reachable nodes. The only attack at the Perception layer refers to sensors connected to the LoRaWan controller. Electromagnetic interference attacks can influence sensory measurements and actuation control.
	
	\textit{$RM_H$~\cite{Hanetal2014}:}
	We encounter privacy and policy-related issues at the Information layer, desynchronization problems at the Supervisory Control sublayer, and issues with direct intervention at the Physical layer. The Information layer is mainly protected by encryption; however, this does not stop attackers from traffic analysis; gathering event-based information such as pedestrian or vehicle passing results are helpful, e.g., to assess citizens' behavioral patterns in their neighborhood. Furthermore, multiple join attempts caused through interference signals may help an attacker de-crypt keys used for network and application sessions through excuse attacks. In addition, an adversary can tamper with sensory devices on the physical layer to manipulate measurements and influence lamp control, e.g., artificially raise sky illumination levels, tricking the system into believing that a shallow street illumination level suffices. Finally, a control issue that might emerge is a side effect of scalability. Similar to the situation described in Section~\ref{sec:dali}, the size of the network influences the throughput capabilities. Even though the control loop involving LoRa is less tight, an extended period of reduced or interrupted communication with a gateway or network server could lead to unpredictable behavior. 	
	
	\textit{$RM_V$~\cite{Vargaetal2017}:} Similar to DALI, configuration tampering attacks at the Application layer may befall LoRaWan end-nodes with similar side effects. Finally, at the Network layer, the model adds fairness mechanism attacks and extends the definition of DoS flooding. The former attack tampers with the open-source WAN algorithm to elude medium sharing mechanisms and exhausting transmission resources. Flooding's extended definition reveals a similar purpose: malformed packets flood a targeted network or application to overload and corrupt resource availability. 
	
	\subsubsection{IP-Based Infrastructure}
	The most traditional network in our SLA, the IP infrastructure, connects multiple IP-based devices. It transports data between the on-site LoRaWan gateways (C) and the IoT computing cloud (F) through dedicated firewalls (D-E), Figure~\ref{fig:aLayout}. In addition, the network is in charge of a higher level of connectivity, servicing LoRaWan and DALI for the applications supervising the city's lighting.	
	
	\textit{$RM_A$~\cite{AshibaniMahmoud2017}:} On the Transmission layer, we find typical communication-related attacks that target resource availability or intercept or manipulate messages. Most parts of the network apply double-encryption, making attacks such as MitM and eavesdropping onerous. Routing-based attacks are most effective on routed LAN packets, available at the Cloud internal LAN (7). Selective forwarding, routing, sinkhole, wormhole, replay, spoofing, or compromised key attacks could occur here. They help an attacker to weaken and delay network traffic or reroute data for traffic and side-channel analysis. If integrated with traffic analysis, such attacks get more efficient and difficult to detect.
	
	Besides, despite tunneling and encryption, most of the DoS attacks keep their effectiveness. A typical DDoS attack could target VPN end-points, e.g., FW (E), which makes up a single point of failure for the two sub-nets, and a bottleneck on high traffic. Similarly, all network components are susceptible to exhaustion attacks. Finally, tampering and node capture, e.g., the external firewall, could help acquire stored secrets and, e.g., reroute VPN tunnels for general data capture. The primary function of the Application layer is storing and elaboration of information. Primary attacks to this layer identify thus as Database attacks, including data alteration and User Privacy leakage through data mining on the sensed data. An attacker may gain access to a system via malicious code on shared instances or buffer overflow and consequent code injection.
	
	Furthermore, along with continuously more service-oriented systems, service discovery spoofing helps integrate malicious services into the system, gathering data access. Replayed messages on this service plane may help an attacker to get the trust of the system. Message interception and alteration (MitM) and eavesdropping can cause data leaks or corruption. A malicious service can flood other services until exhaustion, making them unavailable. Such attacks' effectiveness depends on the architecture and implementation of the data processing cloud, not specified by any examined standard.
	
	\textit{$RM_H$~\cite{Hanetal2014}:} %Although no new attacks are present in the Information layer, 
	The Network layer presents re-definitions of Sybil and spoofing. For example, injected routing error messages at the inter-VM LAN connection (7) make the grid seem partially offline. At the same time, Sybil attacks target fake network size. Finally, on the Control layer, the system keeps being the target of desynchronization attacks. An attacker can, e.g., tamper with time-servers to misalign lamp control from status. 
	
	\textit{$RM_V$~\cite{Vargaetal2017}:} $ RM_V$'s application layer considers user interaction with system and data separately from its computation. Thus, if a user connects remotely to the system, a new path opens, allowing network-based threats similar to $RM_A$, including eavesdropping, MitM, routing, or system exhaustion attacks. Configuration tampering attacks may further affect the new terminal, attempting to influence the lighting system's function remotely. On the Data processing layer, we identify Malware attacks again to gain system-level access. $RM_V$ further highlights the interactions and attacks that might occur inter-VM and based on shared resources' contention. The former include instant-on gap attacks, where immediate demand requirements allow initial unrestrained executions due to performance concerns. The latter rely on the exhaustion of shared resources. As a result, the attacked service is depleted and unable to perform the requested services. Another mentioned attack, exhaustion flooding, achieves a similar result. The flooding with requests requires additional resources, slows down the system, and finally exhausts all resources. Side-channel attacks could extract information from non-sanitized shared memory or CPU caches among the VMs. The model does not include additional attacks for the Networking layer.
	
	\textit{$RM_L$~\cite{Linetal2017-2}:} This model sees user-focused attacks on the application layer during client interaction. They try to leak data and capture user credentials through infected emails, phishing websites, and malicious scripts. The model then adds two more definitions on the Network layer: the sinkhole attack, as a maneuver to get more input data routed through for traffic analysis and device tampering and to secure a device's configuration data and secrets, and consequently, gain unauthorized access to devices and networks.
	
	\subsection{Attack tree and Vulnerabilities}
	\label{sec:vultree}
	
	\subsubsection{Attack tree for Smart-lighting architecture}
	
	\begin{figure}[tb]
		\centering
		\includegraphics[width=\linewidth, height=0.5\textwidth,keepaspectratio]{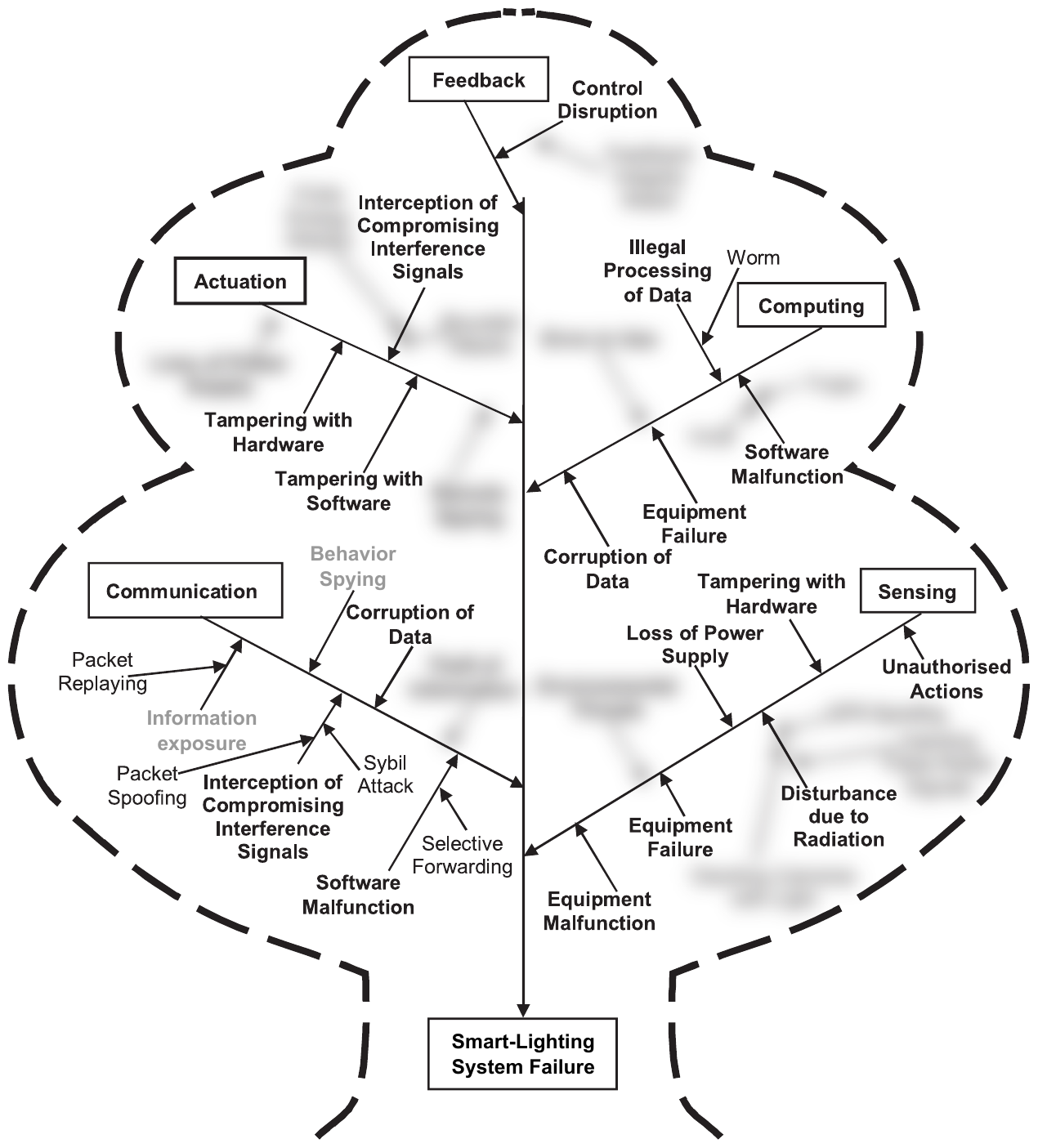}
		\caption{Attack tree for a Smart-Lighting architecture, modified (gray), blurred removed from~\cite{Alguliyevetal2018}.}
		\label{fig:tree}
		\vspace{-10px}	
	\end{figure}	
	
	Based upon the attack and threat tree developed by Alguliyev \textit{et al.}~\cite{Alguliyevetal2018}, we created a version dedicated to SLA attacks. Figure~\ref{fig:tree} illustrates the resulting attacks-threats CPS functional model where the gray highlighting marks alterations w.r.t. the original, i.e., renamed or relocated branches. In the following, we detail threats connected to the attacks.
	
	\noindent \textit{Attacks on actuation.} Our SLA of Figure~\ref{fig:aLayout} contains two actuators: DALI ballasts that control the lamps and LoRaWan timed controllers to manage these ballasts. Both are installed mostly on or near a light pole.
	%
	% Tamper hardware to impede actuation - PP: physical damage to CPS
	A threat of \textit{Tampering with Hardware} results when physical interaction with the node can occlude actuation. An attacker can manipulate a LoRaWan end-node or DALI ballast to take control, disable or extract secrets with device tampering or node destruction attacks.
	%
	% Tamper Software to impede actuation - PP: compromise of information -> change the information to 'bad' state?
	\textit{Tampering with Software} occurs when changes on it make actuation non-functional. For instance, Integrity attacks on a lamp-driving LoRaWan-node can cause incorrect lamp switching times, impeding proper lighting. 
	%
	% Incept destabilizing signals, i.e., false commands - PP: compromise function
	Finally, \textit{Interception of compromising interference signals} refers to actuation instability caused by external intervention on the actuator signals in closed-loop systems. For example, an attacker can destabilize lamp control and manipulating switching behavior through command-control forgery attacks on DALI ballast.
	%
	% Remote spying
	% Loss of power supply -> no battery
	
	\noindent \textit{Attacks on Communication:} 
	The communication infrastructure of our SLA is represented by DALI, LoRaWan, and IP-network infrastructures. These include bridges between DALI, LoRaWan, and IP-based components, i.e., LoRaWan end-nodes and Gateways, the two firewalls, and the Internet. In addition, all connections, except DALI, are encrypted at least once; AES128-CBC for LoRaWan, IPSEC, and HTTPS for IP-based connectivity.
	%
	% Information scouting - passive - PP: compromise of information
	\textit{Information exposure} refers to the threat that allows data gathering on a non-protected communication channel. An attacker can listen in and obtain information on system and encryption passively via an eavesdropping attack on DALI networks or actively through polling via replay attacks on LoRaWan channels.
	%
	% Spy activities - PP: compromise of information
	\textit{Behavior spying} results when an attacker can gather long-term information on the system's operation, people, and activity remotely. Via traffic analysis attacks, e.g., an adversary, can inspect a LoRaWan end-node's event-based transmissions that report pedestrian movement. As stated in Section~\ref{sec:attacks}, such circumstantial information can help determine citizens' whereabouts for planned burglary.
	%
	% This includes resource exhaustion when buffers are full. The packets are dropped due to flooding -- PP: compromise function
	\textit{Software malfunction} results from circumstances that cause incoherent, incomplete, or timely inadequate data transmission that inhibit the system's correct operation. Typical attacks that might cause such behavior are selective forwarding or flooding attacks, applicable to every IP network link. Another example is collision attacks that delay the successful reception of event-based packets from the LoRaWan end-node until a successful re-transmission attempt.
	%
	% Manipulation of data traffic - PP: Unauthorized action 
	The threat of \textit{Corruption of data} occurs when an attacker can manipulate information and thus void data integrity. On DALI networks, e.g., the adversary could easily tamper with the transit data as the protocol has no encryption or access control.
	%
	% Interference in packet delivery, interfering with communication function -- PP: compromise function, forging and abuse of rights
	\textit{Interception of compromising interference signals}, again, refers to communication instability caused by external intervention on the data transmission. Such instability can be caused by jamming attacks on the LoRaWan network or spoofing attacks on the IP connectivity and flooding attack with consequent loss or alteration of packets or connectivity.
	%
	% Extraction of information from data traffic. - PP: compromise of information
	%\textit{Theft of information} on communication occurs when an attacker can inspect data packets and gather private information on users and their possessions. 
	% NO INFORMATION ON USERS AND POSSESSIONS ON THE SYSTEM - DIFFERENT FOR SWATER
	
	\noindent \textit{Attacks on feedback:} 
	Feedback refers to the control function that Cyber-physical systems perform when acting through actuators on sensory input or computational status changes. These include, thus, control algorithms and systems for their implementation.
	%
	% Interference with a control feedback loop -- PP: loss of essential services
	\textit{Control disruption} occurs when the system cannot react to sensory input or status changes, thus destabilizing a system. Via a control-command forgery attack, an attacker could, e.g., manipulate the status of a DALI ballast, desynchronizing feedback control and influencing correct actuation.
	
	\noindent \textit{Attacks on Computing:} 
	Computing refers to the equipment used for data storage and elaboration. Cloud services and infrastructure (F) serve data mining, user interaction, and process performance improvement.
	%
	% Corruption of stored information  -- PP: loss of essential services
	The threat of \textit{Corruption of data} refers to manipulating information, stored and computed values, e.g., programmed light switching times, to secretly damage the system. A data tampering or integrity attack can alter stored control information.
	%	
	% Completion of execution - Availability of service -- PP: technical failure
	The \textit{Equipment failure} occurs when the computing infrastructure is unable to fulfill the requested computation task. These failures can happen due to physical wear-out and resource exhaustion, an attack that depletes computing resources.
	%
	% Correctness of execution -- PP: compromise function,
	\textit{Software malfunction}, yet, results when the computation does execute as requested, but not correctly. These malfunctions are often caused by bugs but can also be due to malicious code installed in the cloud servers, e.g., viruses and trojans, which tamper with software functionality.
	%
	% Unauthorized access and processing of data -- PP: compromise function
	Finally, \textit{Illegal data processing} happens when an unauthorized agent or a user accesses more than the allowed amount of resources and data and discloses user privacy. For example, such exposure can be a consequence of installed malware (Worms) or an attacker that performs side-channel attacks. In addition, a maligned virtual machine on a multi-tenant cloud could tap shared memory and manipulate the computing instance.
	
	\noindent \textit{Attacks on Sensing:} 
	Sensing in our SLA is performed on two locations: DALI ballasts that inform about the lamps' real-time data, and LoRaWan controllers, sometimes battery-powered, that sense the environment, e.g., luminosity or movement sensors.
	%
	% Draining of resources for outage -- PP: loss of essential services
	\textit{Loss of Power Supply} is relevant for devices with reduced energy resources that may suffer from exhaustion and fail service. Battery-powered LoRaWan end-node may experience an outage due to forced repeated transmissions through LoRa jamming attacks that sleep-deprived the node.
	%
	% Actions that disable the sensing's equipment function -- PP: loss of essential services
	\textit{Equipment failure}, yet, refers to the inability of nodes to perform the required task. A node outage attack can put a LoRaWan or DALI node out of order via physical destruction.
	%
	% Tampering with the sensory hardware to impede measurement -- PP: physical damage
	\textit{Tampering with hardware} on sensing identifies issues that might arise when hardware modifications impede correct measurement. Direct physical intervention attacks can cover a lighting sensor, making it inoperable.
	%
	% Unauthorized tampering with the sensor function -- PP: compromise of function, forging, and abuse
	\textit{Unauthorized actions} recall the possibility of prohibited intervention on sensors that access or alter data, misuse the node, or impede its function. The sensing configuration data on the unprotected DALI nodes can be manipulated through data tampering attacks during writes on the bus link, altering measured results.
	%
	% Actions on sensing that impede the execution -- PP: compromise function
	The same attack can also be the source of other threats. \textit{Equipment malfunction} is the result of incorrect sensing due to technical hindrance. Tampering with a sensor's configuration would cause sensing to fail its function.
	%
	% Alteration of physical aspects concerning measured data -- PP: radiation malfunctions - voluntary
	Finally, devices are subject to the \textit{Disturbance due to radiation} when an attacker interferes with the normal sensory function by manipulating the measured physical unit. The LoRaWan node. e.g., can be fooled through a physical direct intervention attack, irradiating the luminosity sensor with a torch.
	
	\subsubsection{Vulnerabilities for Smart-lighting architecture}
	
	After the evaluation of attacks and threats for this SLA, we now identify the causing vulnerabilities. Tracing vulnerability records from the related papers~\cite{AshibaniMahmoud2017, Hanetal2014}, we examine threats and attacks to detect possible vulnerabilities.
	
	At the perception and transmission layers of $ RM_A $, most of the attacks identified have two common causes: the low resource constraint the devices withhold and their physical size and exposure. Resource limitation is mostly the enabler of attacks that hinder proper communication, protection, and access control. Unprotected DALI allows an attacker to eavesdrop or inject any command or data. Targeted LoRa or DALI network attacks can deplete available communication or energy resources, disabling parts of the network and feedback control. Similarly, the limited ether availability constraints the transmission capacity of LoRaWan and eases the attacker's channel interference.
	
	Furthermore, the large scale of an SLA contributes to resource scarcity. It increases channel contention and utilization and co-existence problems~\cite{Ghenaetal2019}, finally forcing air-time management or transmission power throttling to reduce range and interference rate. The Wide distribution of a lighting system conduces to the vulnerability of physical exposure. Unattended areas ease network integrity attacks through device tampering, targeted interference, and device destruction. It makes nodes accessible and allows for physical interaction, altering measurements and feedback. Similarly, on the transmission layer, the SLA's wide distribution and large scale cause LoRa's ether resources to incur bottlenecks if an incorrect device configuration neglects available channels. The same holds for gateway setup where incorrect settings can ease preamble-based resource availability attacks.
	
	Software bugs and inconsistent protocols may enable unauthorized access to infrastructure and information on the transmission and application layer.  Human-made error or incorrect device configuration may allow attackers to access systems due to incorrect or mixed permissions schemes or cause system failure. Further vulnerabilities present at the IP and Cloud infrastructure are mostly the typical issues encountered in modern systems. We find missing specification details for the software components running the Smart-* architecture's back-end in addition to service attacks and information leakage issues. Indeed, the two non-standard components, an IDS (D-E) for CPS and the network server (F), have not been defined thoroughly in their specification and architecture~\cite{Lora2017-2, Hanetal2014}. While we can secure the rest of the IP system by applying traditional architectural patterns and techniques, these two components suffer from inconsistent or incomplete specifications.
	
	\subsection{Reflection on countermeasures}
	This section offers some countermeasures specific to the Smart-Lighting system's weaknesses under study that we leverage from the analysis in the previous sections, the existing literature, and specifications of the technologies. The list should not be seen as exhaustive. 
	
	At the lamp end-posts, we have to deal primarily with the physical exposure of the DALI bus, the LoRaWan controllers, and the sensors. Using cabinets and locks that require a specific tool or key and mounting controllers at height might impede immediate access to wires and devices. Wires should further be carried through shielded conducts, diminishing the risk of interference. Unfortunately, resource constraints and the limiting standards do not permit protection measures against eavesdropping or MitM attacks for DALI communication; a replacement with more powerful hardware could significantly impact unit installation cost and solution attractiveness.
	
	One main point that helps mitigate the attacks on the limited ether availability is a balanced configuration of the LoRaWan network. The LoRaWan standard provides the network server with the ability to reconfigure channels and optimize ether usage for gateways and end-nodes. However, there is no binding requirement for such capability. To our knowledge, no network server product includes neither an automatic channel distribution on gateways and nodes nor the ability to select channels beyond the eight frequencies in the Semtech default profile. 
	
	Proper distribution of bandwidth and frequencies can drastically increase the resilience of the infrastructure. The sixteen-plus available settings-slots prescribed by the LoRaWan standard allow a complementary configuration. Furthermore, each node should reach at least two gateways using the lowest spread factor. This approach increases the communication robustness due to LoRa's high channel selectivity and switchable, more robust, higher symbol rates. The latter makes it easy to increase the allowed SNR range and decode messages despite interferences~\cite{Lora2017, Semtech2015}.
	
	Likewise, adjacent nodes' downlink and uplink settings should be distributed equally among reachable gateways and channels. End-nodes typically communicate on two channels: one randomly selected among the enabled channels for bidirectional transfer and a second shared RX window from the gateway to all nodes. This second link adds resilience to the network. As long as one downlink is available, the network server can reconfigure a node to use new uplink frequencies~\cite{Lora2017}. If possible, node channel configurations should contain a disabled configuration of all gateway channels in reach. Disabled channels are automatically enabled after several unsuccessful transmissions, empowering a node in distress to reach all available gateways. 
	
	An algorithm running on the network server may manage such additional channel configurations to exploit maximum robustness. It might use geo-information and empirical measurement results to compute channel distribution appropriately and send updates over the secondary RX window. An algorithm for this purpose has been developed by Demetri et al.~\cite{Demetrietal2019}. Satellite imaging and experimental measurements approximate signal coverage and considers the environment, locations, buildings, and city structure. To avoid the issue of limited throughput and co-existence interference, the number of nodes per channel and gateway should also be equally distributed~\cite{Augustinetal2016}. A tool called LoRaSim by the university of Lancaster~\footnote{\url{https://www.lancaster.ac.uk/scc/sites/lora/lorasim.html}} helps this purpose. Although the tool does not consider the environmental situation, it can verify if a configuration is viable. It selects optimal frequencies, captures situations of hidden terminals and exposed nodes, and determines the best-case range and coverage for a given network configuration. Lower spread factor and less interference reduce required air-time and repetition. A service incorporating such an algorithm could improve overall resilience, optimize hardware use and increase end-node battery lifetime. 
	
	Unfortunately, LoRa (physical layer) and specification-dependent vulnerabilities cannot directly be dealt with. The specification of protocols is an alliance product (DiiA and LoRa Alliance) and might be open to improvement proposals~\cite{Lora2017}. At the moment, different proposals exist for both vulnerabilities~\cite{Adelantadoetal2017}. The alliance also recently proposed an intra-channel hopping technique (FHSS) to mitigate collisions and contention~\cite{Lora2020}. The higher robustness comes at a price of a very low throughput of only a few hundred bps. It promises to be an elegant solution for high-density and coexisting networks. However, such changes need time for validation and processing and can therefore be considered only in the long run.
	
	Simple stateful packet inspection is not enough for a CPS's IP-based network. Han \textit{et al.}~\cite{Hanetal2014} identifies security challenges not uniquely at the border to the external networks, but everywhere in this complex interconnected and heterogeneous system. Therefore, intrusion detection must be entwined in the whole CPS system according to each node's limits. Furthermore, each node could be tampered with generating invalid data. Thus, the solution extends from brute physical force and consequent failures to uncertain information degraded and influencing a system's control. Finally, border firewalls are often the responsible routing point for point-to-point networks. Ideally, multiple connections between IP-based networks can avoid bottlenecks and targeting attacks by routing traffic as needed.
	
	%Misconfiguaration, SW bugs, specification miss!	
	The final set of discussed vulnerabilities connects solely to the application layer. Most of the software modules of the control units and in the application cloud work with parameters. To avoid that those settings are invalid, ideally, the final device or application that uses the information must verify correctness. Han \textit{et al.} see this also as a possible application for an IDS as an adversary could inject invalid values to cause a control deviation or misbehavior. However, the limited resources make a distributed IDS difficult on some devices. Therefore, a parameter check, a distributed IDS, or both should be installed based on resource availability.
	
	More problems arise if the specifications for these software modules have errors or are incomplete. Unfortunately, in this case, the specifications should also follow standards and might suffer from this dependency. Nevertheless, many details can be derived and adapted following best practices and generalizations from experience with similar installations and architectures. Multi-tenant microservice-based systems are popular in cloud-based computation, making them an excellent architectural template source. Therefore, a computing cloud infrastructure could be derived from microservice-based architectures for data elaboration, integrated with the knowledge gained from running experiments and prototypes. These should finally help achieve the highest security standards without impacting the overall performance. Lastly, most software is following new technology trends, subject to multiple changes in a short time, and suffering from high defect probability. Therefore, agile practice and testing tool-chains are the only suggestions to be given from the development standpoint. 
	
	\section{Discussion and Conclusions}
	\label{sec:concl}
	
	This security analysis presented a technique for the offline analysis of a Smart-* multi-domain system-of-systems. We proposed a design science approach that relied on the connected domains' experiences and performed a layer-based cross-analysis on a Smart-Lighting use case. Using four distinct layered architecture modeling approaches, we identified architectural roles, assigned model layers. Then, we created a unified taxonomy that reflects and extends each involved domain's attack definitions, threats, and vulnerabilities. In an iterative process, we determined possible attacks, valid threats and discussed vulnerabilities for a merged-domain Smart-Lighting architecture. Finally, we discussed some first possible countermeasures. 
	
	After the execution of our analysis, we can assess three significant discoveries for Industry 4.0. Firstly, the domain overlapping configuration of such a system-of-systems makes it infeasible to cover all threats and attacks based on a single domain's viewpoint. The integrative approach we presented detected more issues than a single model would. Interestingly, we find the most model divergence in the ``cyber''-layers, where computation and decision occur, while most data exchange and physical interaction layers remain unchanged. This consistency is probably because gathering, actuation, and data transport are a joint function of all four analyzed papers. When integrating future analyses with other studies, we expect changes in the upper architecture layers only. Secondly, the changing focus of the discussed models highlights aspects of a heterogeneous system. It proves that the new multi-domain architecture inherits many, if not all, characteristics of the involved domains. For example, Cloud-security issues are not a typical concern for traditional control-oriented CPS. Thirdly, vulnerabilities, threats, and attacks may alter definition, range, and weight depending on the application domain. The taxonomy table and Section~\ref{sec:vultree} show how similar threat or attack names can have different definitions and applications that the domain of origin might influence. It is thus reasonable to pre-define and clarify all taxonomy before reaching conclusions. However, the resulting multi-domain taxonomy is a product of the involved reference models' role, layer, and attack allocation. Each new system-of-systems analysis requires thus repeating or refining the present analysis.
	
	Future work will test and extend the results of this analysis. Through a second study case, we will analyze the change and variability of detected issues. Simultaneously, on-site tests will help validate the extent and risks of the vulnerabilities involved.
	
	\section*{Acknowledgments}
	We thank Systems S.r.l for the funding and support of this project under the ``Industry 4.0 for the Smart-* (I4S)'' project.
	
	\bibliographystyle{ACM-Reference-Format}
	\bibliography{papers}
	
\end{document}